\documentclass[aps,pre,amsmath,showpacs]{revtex4}
\usepackage{graphics}
\usepackage{epsfig}
\begin{document}
\title{Embedded solitons in the third-order nonlinear Schr\"{o}dinger equation }
\author{Debabrata Pal, Sk. Golam Ali }
\author{B. Talukdar}
\email{binoyt123@rediffmail.com}
\affiliation{Department of Physics, Visva-Bharati University,
Santiniketan 731235, India}
\begin{abstract}
We derive a straightforward variational method to construct embedded soliton solutions of the third-order nonlinear Sch\"odinger equation and analytically demonstrate that these solitons exist as a continuous family.
We argue that a particular embedded soliton when perturbed may always relax to the adjacent one so as to make it fully stable. 
\end{abstract}
\maketitle

\section{ Introduction}
The distortion of pulses propagating through  waveguides arises due to dispersive spreading. However, if the waveguides is made using materials the refractive index of which depends nonlinearly on the intensity of the propagating pulses, it will be possible to compensate the effect of dispersion by a pulse-narrowing effect \cite{1}. A typical example in respect of this is provided by optical fibers. Here the weak nonlinearity
of the index of refraction arising due to Kerr effect produces a self modulation which in turn causes steepening of the wave. The dynamical interplay between the dispersive and narrowing effects produces the so called optical solitons. For an isotropic medium with cubic nonlinearity the complex optical wave field $\phi(x,t)$ is governed by the nonlinear Schr\"{o}dinger equation (NLS)

\begin{equation}
i\phi_x+\phi_{2t}+|\phi|^2\phi=0, \,\,\,\, \phi=\phi(x,t).
\end{equation}
Here $t$ stands for time and $x$, the coordinate along the fiber. The suffixes $x$ and $t$ of $\phi$
denote partial derivatives with respect to these variables. In particular, $\phi_{2t}=\dfrac{\partial^2\phi}{\partial t^2}$.
The main reason for the robustness of the solitons described by $(1)$ is that the wave numbers of the solitons lie in a range that is forbidden for the linear dispersive waves \cite{2}. As a result, linear wave 
can not be in resonance with the soliton to receive energy from the latter. We then have a stable soliton.
\par 
The propagation of picosecond pulses is well described by the NLS equation in $(1)$ which accounts for only the second-order dispersion and self modulation. But for femtosecond pulses another important physical effect namely, the third-order dispersion comes into play. In this case the appropriate evolution equation for the pulse propagation is given by \cite{3}
\begin{equation}
i\phi_x+\phi_{2t}+|\phi|^2\phi=i\beta\phi_{3t}.
\end{equation}
Understandably, the parameter $\beta$ is a measure of the perturbation caused by the third-order dispersion. Due to the effect of perturbation a soliton could have wave numbers residing inside the linear spectrum of the system. Such a soliton is called an embedded soliton \cite{4}. Clearly, the embedded solitons can be in resonance with the perturbation to generate continuous wave radiation for subsequent decay. Despite this, in certain parameter regimes these solitons can be stable \cite{5}. In view of this the stability analysis of embedded solitons arising in the solutions of higher-order NLS equation has been subject of considerable current interest. The present work is an effort in this direction. 
\par
Studies in embedded solitons \cite{6}  within the framework of a third-order NLS equation indicate that these soliton exits as a continuous family. The members of the family being characterized by their propagation velocities are connected through a gauge transformation \cite{3}. We demonstrate the existance of such a family
 by using a straightforward analytical model and argue that embedded solitons can be fully stable despite their tendency to shed radiation. We believe 
that such an approach has distinct advantages to deal with physical problems because many unknown effects are then 
readily expressed and evaluated. To achieve our goal we shall derive a variational approach involving trial functions in order to describe the main characteristics of the embedded solitons as determined by the third-order 
NLS equation in $(2)$. Although the chosen trial functions have a specific form , the shape parameters are allowed to evolve as the solitons propagate. With this assumtion the evolution equation under consideration simplifies to a reduced Lagrangian problem \cite{7}. We shall try to study the stability of embedded 
solitons  
by taking variations of this Lagrangian with respect to the shape parameters.

\section{ Variational description of the third-order NLS equation}
For the NLS equation in $(1)$ there exists a well-defined spectral problem \cite{8} such that one can write a closed
form analytical solution of it in terms of sech functions. As opposed to this, the third-order NLS equation can not be solved
analytically. Keeping this in view we look for a variational treatment of $(2)$ to find an accurate approximation solution. For our treatment we will essentially rely on a Ritz optimization procedure \cite{9}
 applied to the Lagrangian function for the NLS equation in $(2)$. We have found that the action functional
\begin{equation}
W=\int\int{\cal L}\left( \phi,\phi^{\star},\phi_x,\phi_x^{\star},\phi_t,\phi_{2t},\phi_t^{\star},\phi_{2t}^{\star}\right)  dxdt
\end{equation} 
with the Lagrangian density
\begin{equation}
{\cal L}=\frac{1}{2}\left( \phi\phi_x^{\star}-\phi^{\star}\phi_x\right)-\frac{1}{2}\phi^2{\phi^{\star}}^{2}
+\phi_t\phi_t^{\star}+\frac{i}{2}\beta\left( \phi_t\phi_{2t}^{\star}-\phi_t^{\star}\phi_{2t}\right) 
\end{equation} 
reproduces the third-order NLS equation via the Hamilton's variational principle.
\par
In the Ritz optimization procedure, the first variation of the action functional is made to vanish for a suitable
chosen trial function. In analogy with the well-known solution of $(1)$ we introduce a sech type trial function
\begin{eqnarray}
\phi(x,t)=\eta(x) \text{sech}\left[ \frac{t-y(x)}{a(x)} \right]\times \,\,\,\,\,\,\nonumber \\\exp i\left[ V(x)(t-y(x))+\frac{b(x)}{2a(x)}\left( t-y(x)\right)^2+\sigma(x) \right]  .
\end{eqnarray} 
Here $\eta$, $y$, $V$, $a$, $b$, and $\sigma$ are real functions of $x$. The parameters $\eta$, $y$ and $a$ are related to the three lowest-order moments of the $\phi$ envelope and represent
respectively its amplitude, central position and width respectively. The other parameters $\sigma$, $V$ and $b$ stand for the phase, velocity (centre of the soliton)  and frequency chirp. Understandably, these parameters will all vary with the distance of propagation. Using $(5)$ in $(4)$ we get
\begin{equation}
{\cal L}_s=\sum_{i=1}^{4}{\cal L}^{(i)}_{s},
\end{equation}
where

\begin{widetext}
\begin{subequations}
\begin{equation}
{\cal L}_s^{(1)}=\eta^{2}\left\lbrace a\frac{dV}{dx}\left( \frac{t-y}{a}\right)-V\frac{dy}{dx}+\frac{1}{2}\left( a\frac{db}{dx}-b\frac{da}{dx}\right) \left( \frac{t-y}{a}\right)^2-b\frac{dy}{dx} \left(\frac{t-y}{a}\right)+\frac{d\sigma}{dx} \right\rbrace sech^2\left(\frac{t-y}{a}\right),
\end{equation}
\begin{equation}
 {\cal L}_s^{(2)}= -\frac{1}{2}\eta^{4}sech^4\left(\frac{t-y}{a}\right),
\end{equation}
\begin{equation}
{\cal L}_s^{(3)}=\eta^{2}\left[ \frac{1}{a^2}tanh^2\left(\frac{t-y}{a}\right)+ \left\lbrace V+\frac{b(t-y)}{a}\right\rbrace ^2\right]sech^2\left(\frac{t-y}{a}\right) 
\end{equation}
and
\begin{eqnarray}
{\cal L}_s^{(4)}=-\beta\eta^{2}{\bigg\lgroup} \frac{b}{a^2}tanh\left(\frac{t-y}{a}\right)-\frac{1}{a^2}\left\lbrace V+\frac{b(t-y)}{a}\right\rbrace tanh^2\left(\frac{t-y}{a}\right)\nonumber\\-\frac{1}{a^2}\left\lbrace V+\frac{b(t-y)}{a}\right\rbrace sech^2\left(\frac{t-y}{a}\right)-\left\lbrace V+\frac{b(t-y)}{a}\right\rbrace^3 {\bigg\rgroup}sech^2\left(\frac{t-y}{a}\right) .
\end{eqnarray}
\end{subequations}
\end{widetext}
Here the subscript $s$ on ${\cal L}$ indicates that we have inserted the $sech$ ansatz for $\phi(x,t)$ into the Lagrangian density. In terms of $(6)$ the variational principle implied by $(3)$ leads to 
\begin{equation}
\delta\int\left\langle L\right\rangle  dx=0
\end{equation} 
with
\begin{equation}
\left\langle L\right\rangle =\int_{-\infty}^{+\infty}{\cal L}_s dt.
\end{equation} 
We have found that result for $\left\langle L\right\rangle$ is given by 
\begin{widetext}
\begin{eqnarray}
\left\langle L\right\rangle=a\eta^{2}{\bigg\lgroup}-2V\frac{dy}{dx}+\frac{\pi^2}{12}\left( a\frac{db}{dx}-b\frac{da}{dx}\right) +\frac{d\sigma}{dx}-\frac{2}{3}\eta^2+\frac{2}{3a^2}+2V^2+\frac{\pi^2b^2}{6}+\beta V\left( \frac{2}{a^2}+2V^2+\frac{\pi^2b^2}{2}\right) {\bigg\rgroup}.
\end{eqnarray} 
\end{widetext}
\section{ Parameters of the trial function}
From the vanishing conditions of the variationals $\frac{\delta \left\langle L\right\rangle}{\delta \sigma}$, 
$\frac{\delta \left\langle L\right\rangle}{\delta \eta}$, $\frac{\delta \left\langle L\right\rangle}{\delta y}$,
$\frac{\delta \left\langle L\right\rangle}{\delta a}$, $\frac{\delta \left\langle L\right\rangle}{\delta b}$ and
$\frac{\delta \left\langle L\right\rangle}{\delta V}$
we obtain the following equations.
\begin{widetext}
\begin{subequations}
\begin{equation}
\frac{d}{dx}\left(a\eta^2 \right)=0,
\end{equation}
\begin{eqnarray}
-2V\frac{dy}{dx}+\frac{\pi^2}{12}\left( a\frac{db}{dx}-b\frac{da}{dx}\right) +\frac{d\sigma}{dx}-\frac{4}{3}\eta^2+\frac{2}{3a^2}+2V^2+\frac{\pi^2b^2}{6}+ \beta V\left( \frac{2}{a^2}+2V^2+\frac{\pi^2b^2}{2}\right)=0,
\end{eqnarray}
\begin{equation}
\frac{d}{dx}\left( 2a\eta^2V\right) =0 ,
\end{equation}  
\begin{eqnarray}
-2V\frac{dy}{dx}+\frac{\pi^2}{12}\left( a\frac{db}{dx}-b\frac{da}{dx}\right) +\frac{d\sigma}{dx}-\frac{2}{3}\eta^2+\frac{2}{3a^2}+2V^2+\frac{\pi^2b^2}{6}+ \nonumber  \\ \beta V\left( \frac{2}{a^2}+2V^2+ \frac{\pi^2b^2}{2}\right) +a\left( \frac{\pi^2}{12}\frac{db}{dx}-\frac{4}{3a^3}-\frac{4\beta V}{a^3}\right) +\frac{\pi^2}{12\eta^2}\frac{d}{dx}\left( a\eta^2b\right) =0,
\end{eqnarray}
\begin{equation}
a\left( -\frac{1}{12}\frac{da}{dx}+\frac{b}{3}+\beta Vb\right)- \frac{1}{12\eta^2}\frac{d}{dx}\left( a^2\eta^2\right) =0
\end{equation} 
and
\begin{equation}
\frac{dy}{dx}-2V-\beta\left( \frac{1}{a^2}+3V^2+\frac{\pi^2b^2}{4}\right) =0.
\end{equation} 
\end{subequations}
From $(11a)$
\begin{subequations}
\begin{equation}
a\eta^2={\rm{constant}}={\rm{E}}_0 ({\rm{say}}) .
\end{equation} 
Use of $(12a)$ in $(11c)$ gives 
\begin{equation}
V={\rm{constant}} .
\end{equation} 
Again $(12a)$ and $(11d)$ give 
\begin{eqnarray}
-2V\frac{dy}{dx}+\frac{\pi^2a}{4}\frac{db}{dx}-\frac{\pi^2b}{12}\frac{da}{dx}+\frac{d\sigma}{dx}-\frac{2}{3}\eta^2-\frac{2}{3}a^2+2V^2+\frac{\pi^2b^2}{6}+\beta V\left( 2V^2+\frac{\pi^2b^2}{2}-\frac{2}{a^2}\right) =0
\end{eqnarray}
Similarly from $(12a)$ and $(11e)$ we have
\begin{equation}
\frac{da}{dx}=r b,\,\,r=2+6\beta V.
\end{equation}
\end{subequations}
\end{widetext}
Subtracting $(11b)$ from  $(12c)$ we write
\begin{equation}
\frac{db}{dx}=\frac{4}{\pi^2a}\left( \frac{r}{a^2}-\eta^2\right)  . 
\end{equation} 
From $(12a)$, $(12d)$ and $(13)$ we get a second-order differtial equation for the pulse width
\begin{equation}
\frac{d^2a}{dx^2}=\frac{4r}{\pi^2}\left( \frac{r}{a^3}-\frac{{\rm{E}}_0}{a^2}\right) .
\end{equation}
The first integral of $(14)$ gives 
\begin{equation}
\frac{1}{2}\left( \frac{da}{dx}\right)^{2}+\Pi(a)=0 
\end{equation}
where the potential field
\begin{equation}
\Pi(a)=\frac{2r^2}{\pi^2a^2}-\frac{4r{\rm{E}}_0}{\pi^2a}+K.
\end{equation}
The constant of integration $K$ is determined by the initial conditions of $(14)$.
\section{Potential function formulation for the dynamics of embedded solitons}

First from the initial conditions of $(14)$ written as $a(x)|_{x=0}=a_0$ and $\left( \frac{da}{dx}\right)|_{x=0}=0 $,
we have 
\begin{equation}
K=-\frac{2r^2}{\pi^2a_{0}^2}+\frac{4r{\rm{E}}_0}{\pi^2a_0}.
\end{equation} 
Using $(17)$ in $(16)$ we write 
\begin{equation}
\Pi(a)=\frac{2r^2}{\pi^2a^2}\left(\frac{a^2_{0}}{a^2}-1\right) -\frac{4r{\rm{E}}_0}{\pi^2a}\left(\frac{a_{0}}{a}-1\right) .
\end{equation} 
We now introduce a normalized pulse width $z(x)=\frac{a(x)}{a_{0}}$ to rewrite $(15)$ as 

\begin{equation}
\frac{1}{2}\left( \frac{dz}{dx}\right)^{2}+\Pi(z)=0 
\end{equation}
 with 

\begin{equation}
\Pi(z)=\left( 1+3\beta V\right) \left( \frac{1}{z}-1\right) \left\lbrace \mu\left( 1+3\beta V\right)\left( \frac{1}{z}+1\right) +\nu\right\rbrace 
\end{equation} 
and

\begin{equation}
\mu=\frac{8}{\pi^2a_{0}^4}\,\,\,\,\,\nu=-\frac{8{\rm{E}}_0}{\pi^2a_{0}^2} .
\end{equation} 
The potential function $\Pi(z)$ vanishes for two values of $z$, namely, $z_1=1$ and $z_2=-\frac{\mu \left( 1+3\beta V\right) }{\nu+\mu \left( 1+3\beta V\right)}$
implying that there is a minimum in between these points. The point at which $\Pi(z)$ is minimum can be obtained
 from $\frac{d\Pi(z)}{dz}=0$ as 
\begin{equation}
z_m=-\frac{2\mu \left( 1+3\beta V\right)}{\nu}.
\end{equation} 
In the limit when $\frac{\nu}{\mu}=-2\left( 1+3\beta V\right)$ we have $z_2=1=z_m$ and $\frac{d\Pi(z)}{dz}|_{z=1}=\Pi(z_m)=0$. In this case the potential well degenerates into a single point such that a particle released at that point will stay there. In the context of this work, this signifies that a wave pulse
for which $\frac{\nu}{\mu}=-2\left( 1+3\beta V\right)$ propagates with unchanged shape as a cosequence of  exact balance between nonlinear and dispersive effects. Interestingly, we  now have a continuous family of soliton solutions because  $\frac{\nu}{\mu}$  depends on the continuous  variable $\beta V$. This implies that in the neighbourhood of each embedded soliton there is another one having a slightly higher or lower energy. A particular embedded soliton when perturbed may always relax to the adjacent one. This makes an embedded  soliton fully stable.

\begin{figure}
\includegraphics[width=0.70\columnwidth, angle=-90] {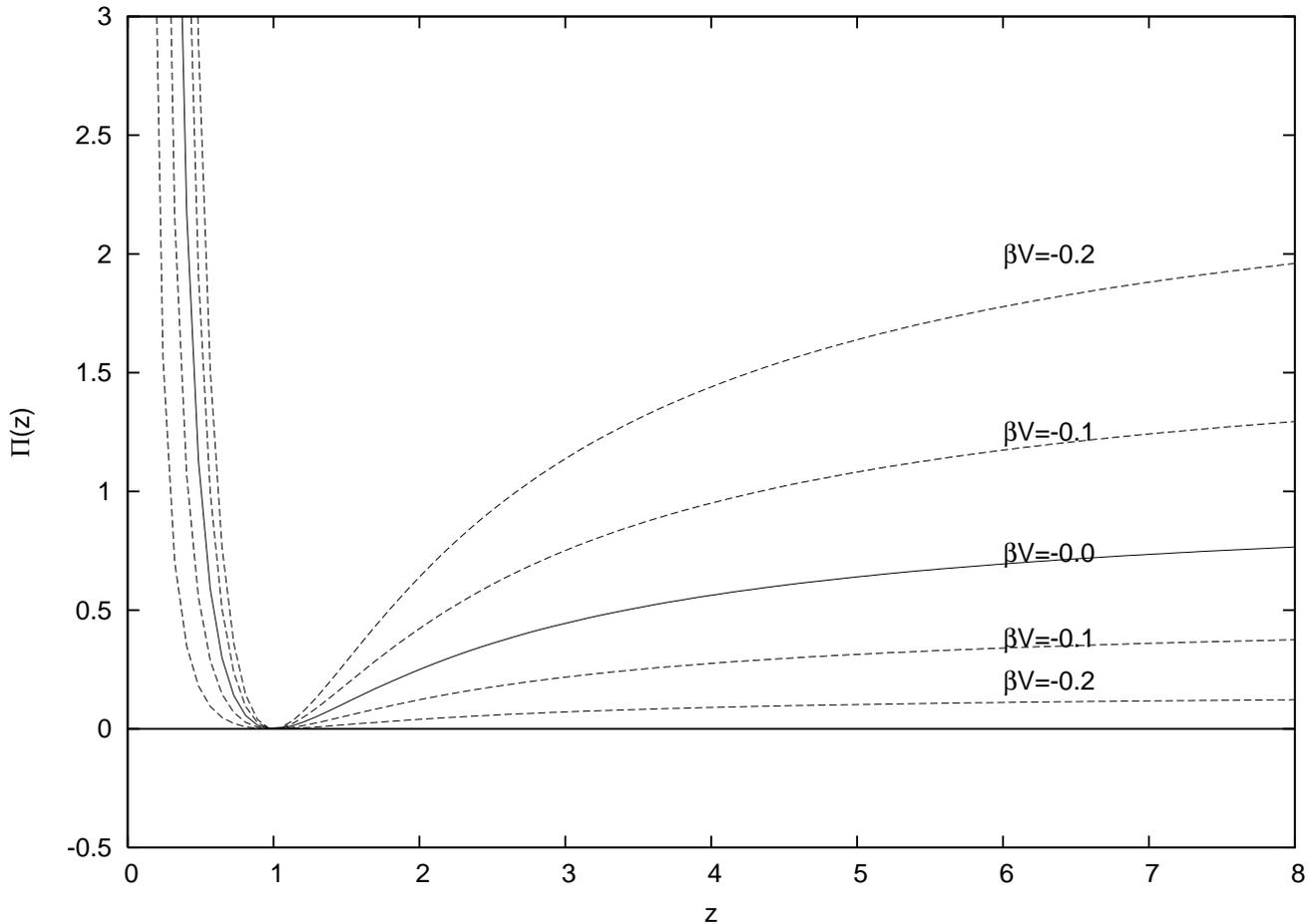}
\caption{${\Pi}(z)$ as a function of $z$ for $\frac{\nu}{\mu}=-2\left( 1+3\beta V\right)$}
\end{figure}

\par
In Fig. 1 we plot the potential function $\Pi(z)$ as a function of $z$ for $\beta V= -0.2, -0.1, 0, 0.1, 0.2$.
Each of these potentials  has $z_1=z_2=z_m=1$ and is, therefore, associated with an embedded soliton. The solid curve $\beta V=0$ is the potential function corresponding to the usual second-order NLS equation while the 
dotted curves represent the potential functions of the embedded solitons as found in the solution of third-order NLS equation. For $\beta V >0$, the potential curves for the embedded solitons lie above the solid curve for $z >1$. For $\beta V <0$ the dotted curves fall below the solid curve. We have drawn only a few curves for discrete values of $\beta V $. However, a dense set of curves can be drwan by varying $\beta V $ continuously.

\section*{ACKNOWLEDGEMENTS}
This  work forms the  part  of a Research Project F.10-10/2003(SR) supported  by the University Grants Commission, 
Govt. of India. One of the authors (SGA) is thankful to the UGC, Govt. of India for a Research Fellowship.

\end{document}